\documentclass[prb,aps, preprint,superscriptaddress,preprintnumbers, endfloats,
]{revtex4-1}
\usepackage{graphicx}
\usepackage{dcolumn}
\usepackage{bm}
\usepackage{color}
\usepackage{sidecap}
\usepackage{amssymb}

\usepackage{tabularx,ragged2e,booktabs}
\usepackage{multirow}

\usepackage{amsmath}
\usepackage{amsfonts}

\begin{document}

\title {Near perfect mode overlap between independently seeded, gain-switched lasers}
\author{L. C. Comandar}
\affiliation{Toshiba Research Europe Limited, 208 Cambridge Science Park, Milton Road, Cambridge, CB4~0GZ, UK}
\affiliation{Cambridge University Engineering Department, 9 JJ Thomson Ave., Cambridge, CB3~0FA, UK}
\author{M. Lucamarini}
\author {B. Fr\"ohlich}
\author {J. F.~Dynes}
\author{Z.~L.~Yuan}
\email{zhiliang.yuan@crl.toshiba.co.uk}
\affiliation{Toshiba Research Europe Limited, 208 Cambridge Science Park, Milton Road, Cambridge, CB4~0GZ, UK}
\author {A.~J.~Shields}
\affiliation{Toshiba Research Europe Limited, 208 Cambridge Science Park, Milton Road, Cambridge, CB4~0GZ, UK}
\date{\today}

\begin{abstract}
We drastically improve the mode overlap between independently seeded, gain-switched laser diodes operating at gigahertz repetition rates by implementing a pulsed light seeding technique. Injecting pulsed light reduces the emission time jitter and enables frequency chirp synchronization while maintaining random optical phases of the emitted laser pulses.
We measure interference of these pulsed sources both in the macroscopic regime, where we demonstrate near perfect mode overlap, and in the single photon regime, where we achieve a Hong-Ou-Mandel dip visibility of $0.499\pm0.004$, thus saturating the theoretical limit of 0.5. The measurement results are reproduced by Monte-Carlo simulations with no free parameters.
Our light source is an ideal solution for generation of high rate, indistinguishable coherent pulses for quantum information applications.
\end{abstract}

\maketitle

Interference between single photons forms the basis for many important quantum information applications, including quantum teleportation,\cite{teleportation1993} quantum repeater,\cite{briegel98} linear optics quantum computing,\cite{knill01} and detector-safe quantum cryptography.\cite{braustein12,lo12} However, any mode mismatch between these photons degrades the interference visibility and subsequently the performance of the applications they are intended for.  The visibility is usually measured with a Hong-Ou-Mandel (HOM) interferometer in the single photon regime.\cite{hong87}
The two photons are sent to interfere on a 50/50 beam-splitter. When they are indistinguishable in all degrees of freedom, they can only emerge from the same output port, as quantum mechanics rules out the possibility of them exiting from different ports.

\begin{figure}[b]
\centering
\includegraphics[width=\columnwidth]{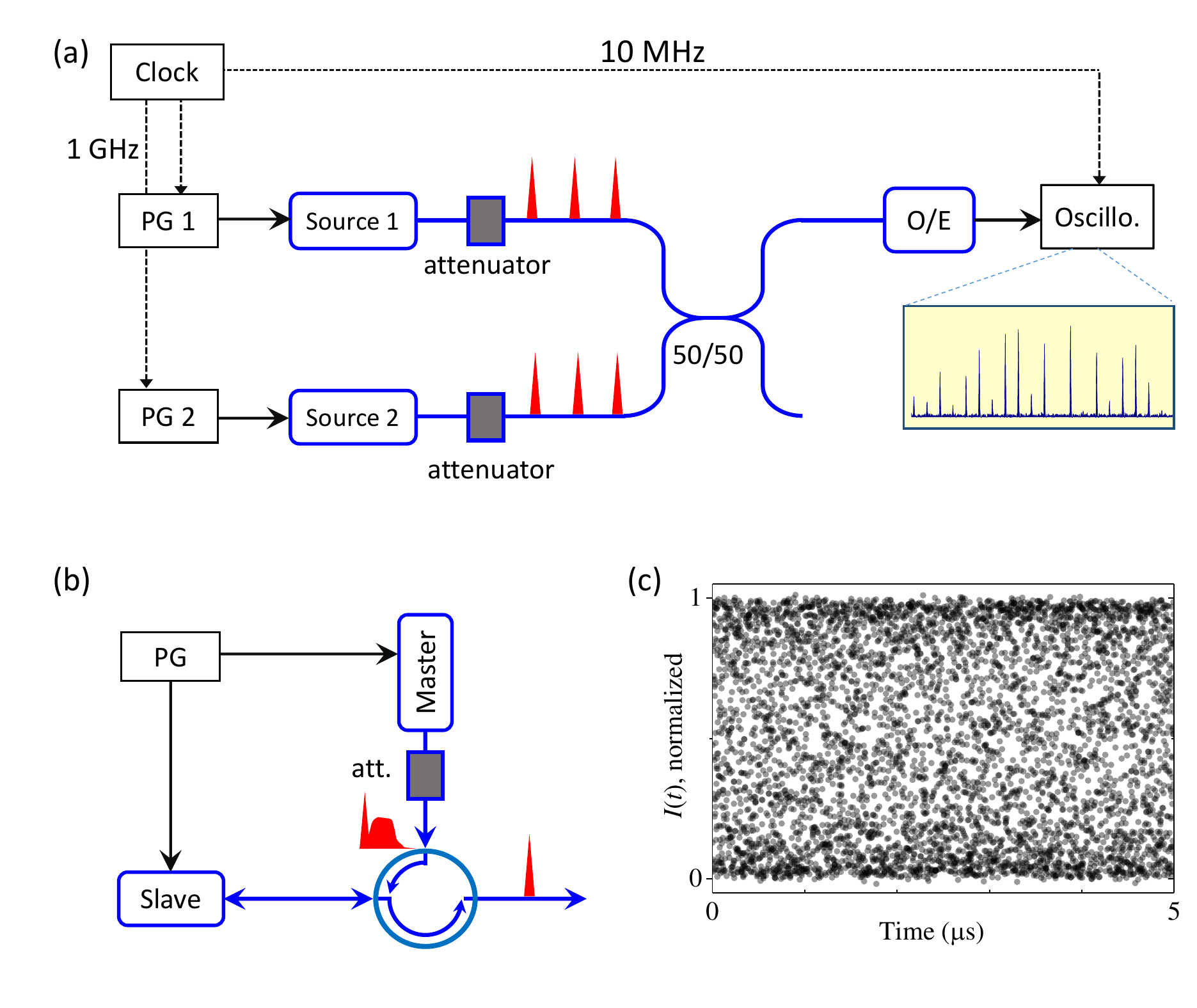}
\caption{(a) Experimental setup for rapid characterization of interference between independent pulsed light sources. The inset shows an interference output trace recorded by the oscilloscope at a sampling rate of 40~GSa/s.
(b) Pulsed light seeding setup. Each light source consists of a pair of gain-switched laser diodes connected via an optical circulator.  The master laser produces a phase-randomized pulse that is used to seed a short pulse from the slave laser.  The seeding optical power is controlled by an attenuator. (c) A 5-$\mu$s data sample of the interference outcomes between two independently seeded, gain switched lasers with the oscilloscope sampling rate set to 1~GHz.
All laser diodes shown here are electrically driven by a 1~GHz square wave voltages superimposed on a DC bias.  O/E: Optical to Electrical Convertor; PG: Pulse Generator.}
\label{fig:fig1}
\newpage
\end{figure}

Attenuated coherent laser pulses can be used to approximate a quantum source in quantum key distribution (QKD) and other forms of quantum communications.\cite{gisin02} In many applications these coherent pulses need to have a high level of indistinguishability.
For the measurement-device-independent (MDI) QKD protocol,\cite{lo12} which relies on two-photon interference to gain immunity from detector vulnerabilities, independent lasers are required to transmit light pulses that can interfere with high visibility.
The required visibility is usually ensured by carving continuous-wave or steady-state laser emission which features narrow spectral linewidth and long coherence time.\cite{silva13,liu2013experimental,rubenok13,tang2014experimental,wang15,tang16} However, the necessity for an intensity modulator and its driving electronics adds to system complexity and cost.
As an additional drawback, a phase modulator is often necessary to randomize the optical phase\cite{liu2013experimental, rubenok13,silva13,tang2014experimental,wang15} in order to meet this condition for the protocol security.

\begin{table}[]
\centering
\caption{Temporal characteristics of the gain-switched slave lasers with or without optical injection.  Full width at half maximum (FWHM) values are used, and we measure these values with an 80~GHz bandwidth sampling oscilloscope which has an intrinsic time jitter of 2.5~ps.}
\label{my-label}
\begin{tabular}[c]{lllll}
\hline \hline
\multicolumn{1}{c|}{} & \multicolumn{2}{c|}{source 1}   & \multicolumn{2}{c}{source 2}      \\ \cline{2-5}
\multicolumn{1}{c|}{} & \multicolumn{1}{c|}{\begin{tabular}[c]{c}width (ps)\end{tabular}} & \multicolumn{1}{c|}{\begin{tabular}[c]{c} jitter (ps)\end{tabular}} & \multicolumn{1}{c|}{\begin{tabular}[c]{c} width (ps)\end{tabular}} & \multicolumn{1}{c}{\begin{tabular}[c]{c}jitter (ps)\end{tabular}} \\ \hline \hline
\multicolumn{1}{c|}{w/o injection} & \multicolumn{1}{c|}{26.5}    & \multicolumn{1}{c|}{10.6}       & \multicolumn{1}{c|}{27.5}    & \multicolumn{1}{c}{12.9}\\ \hline
\multicolumn{1}{c|}{with injection}  & \multicolumn{1}{c|}{28.2}   & \multicolumn{1}{c|}{3.6}       & \multicolumn{1}{c|}{31.8}   & \multicolumn{1}{c}{3.8} \\ \hline \hline

\end{tabular}
\end{table}

Semiconductor laser diodes can be gain-switched to produce naturally phase-randomized, short light pulses at gigahertz clock rate,\cite{yuan14} but these light pulses do not interfere with sufficient visibility required by applications such as MDI-QKD.
Recently, we derived a theoretical model that has identified the cause for their poor interference as the coexistence of frequency chirp and time jitter and we provided experimental verifications through controlling these parameters\cite{yuan2014interference,comandar2015quantum} with the Hong-Ou-Mandel (HOM) dip visibility improved to 0.482.\cite{comandar2015quantum}
Here, we report a further, drastic improvement in the mode overlap between gain-switched laser diodes operating at gigahertz repetition rates.  We measure interference of these pulsed sources in the macroscopic regime, which allows a rapid evaluation optimization of the mode overlap. For independently seeded, gain-switched laser diodes, we achieve a HOM dip visibility of $0.499\pm0.004$, saturating the theoretical limit of 0.5. Our measurement results are reproduced by Monte-Carlo simulations without any free parameter.

\begin{figure}[h]
\centering
\includegraphics[width=\columnwidth]{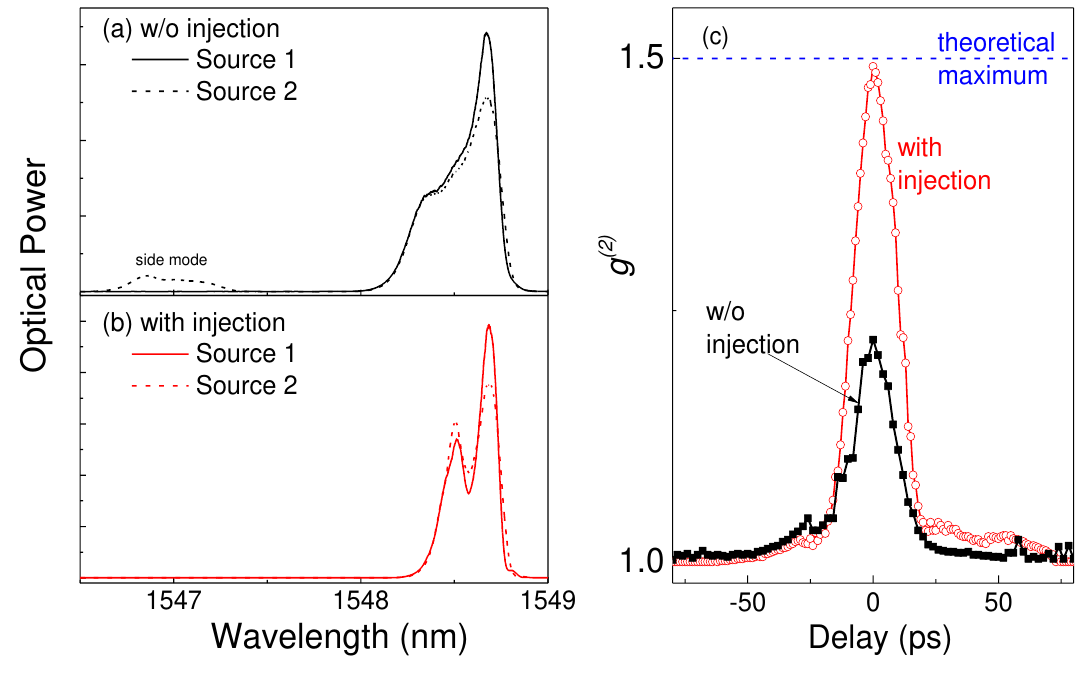}
\caption{The effect of optical injection. (a) The spectra of the slave lasers in the absence of optical injection. The slave 2 shows a strong side mode at the wavelength of 1547~nm. (b) The spectra of the slave lasers in the presence of optical injection. Each master laser is temperature-tuned to allow its wavelength to be resonant with its slave laser.  (c) Second-order intensity correlation ($g^{(2)}$) of the interference output as a function of the time delay between the interfering light sources. The dashed line shows the theoretical maximum that can be obtained with perfectly indistinguishable coherent pulsed sources. }
\label{fig:fig4}
\newpage
\end{figure}

Figure~1(a) shows our experimental setup. We use a signal generator to provide a 1~GHz clock to synchronize two pulse generators (PGs) which drive two independent light sources to produce optical pulses with a gigahertz repetition rate.  These pulses are directed into separate inputs of a fiber-optic beam-splitter with a nominal power splitting ratio of 50/50 (measured ratio: 50.7/49.3), and one of the output ports is fed into an oscilloscope (16 GHz bandwidth) via an optical-electrical convertor.
Both PGs have built-in electronic delay lines with a step size of 1~ps, and we use the delay line in PG2 to adjust the temporal alignment at the 50/50 beam-splitter. The optical intensities are matched using variable optical attenuators. Polarization maintaining (PM) fiber is used to ensure the alignment in the polarization degree of freedom. 

Each light source is formed by a pair of PM fiber-pigtailed distributed feedback (DFB) laser diodes (2.5~GHz bandwidth) connected via a PM optical circulator, see Fig. 1(b).  Two separate master laser diodes are used. Sources 1 and 2 are optically independent and the only link between them is their electronic synchronisation through the common 1~GHz clock (Fig.~1(a)).
All four laser diodes are gain-switched by square wave voltages superimposed on DC bias to transmit 1~GHz optical pulses.  The temporal alignment between a master and its slave is achieved through either the internal PG delays or using electrical signal cables of appropriate lengths.  We use built-in coolers in the laser diodes to temperature-tune their emission wavelengths for spectral alignment.

The oscilloscope shown in Fig.~1(a) is also synchronized to the signal generator via a 10~MHz reference to digitize the interference output at the signal rate of 1~GHz. It has 8-bit vertical amplitude resolution.  A few microseconds acquisition time is sufficient for the built-up of a sequence of the interference outputs $I(t)$ (see Fig. 1(c) as an example) that allows a reliable calculation of the second-order intensity correlation
\begin{equation}\label{eq:xxx}
g^{(2)} = \frac {\langle I^2(t) \rangle} {\langle I(t) \rangle^2},
\end{equation}
\noindent where $\langle X \rangle$ denotes the expectation value of the variable $X$ and the sampling time $t$ is at intervals of 1~ns.
This second-order correlation can be used to quantify the mode match in all possible degrees of freedom
between the light sources.
Under the condition that both light sources produce pulses of a fixed intensity, it has a direct relationship with the HOM dip visibility\cite{rarity05} ($V_{HOM}$) that is measured across the opposite outputs of the interfering beam-splitter,
\begin{equation}
V_{HOM} = g^{(2)} - 1.
\end{equation}
\noindent For perfectly indistinguishable, phase-randomised coherent pulses we expect a theoretical maximum value of $g^{(2)} = 1.5$ ($V_{HOM} = 0.5$).  Any deviation from this extreme suggests distinguishability and degraded interference visibility between the light sources. Coherent pulses of orthogonal modes do not interfere and result in $g^{(2)} = 1$ and $V_{HOM}=0$.

We first characterize the slave lasers in the absence of optical injection. Gain-switched at 1~GHz clock rate, both slave lasers emit pulses with a duration of around 27~ps and a full width half maximum (FWHM) time jitter greater than 10~ps, as summarized in Table I.  We note that in this case each pulse is triggered by spontaneous emission, the random nature of which dominates the measured time jitter.  
Figure~2(a) shows the respective spectra of the lasers after spectrally aligning their emission wavelengths through temperature tuning.  The irregular, broad spectral lineshapes are a result of frequency chirp arising from the carrier density variation in the laser active medium during pulse emission.\cite{yuan14}

In Fig.~2(c) we plot the second-order correlation of the interference output, computed from the oscilloscope acquisition, as a function of temporal overlap between the two light sources.
For long delays, the laser pulses do not overlap in the time domain to interfere, giving rise to a unity correlation.
At zero delay or maximal overlap, the interference between the sources gives a peak correlation of 1.23, which is far below the theoretical maximum of 1.5. This implies distinguishability of the light sources, and we attribute it to the time jitter of the pulsed emissions. As shown in Table I, the measured time jitter amounts to $\sim$40\% of their respective pulse widths. It not only prevents an optimal temporal overlap,
but - more detrimentally - causes a significant amount of dephasing between the interfering sources due to the coexistence of frequency chirp.\cite{yuan14,yuan2014interference} Considering two wave-packets with identical frequency profile $\nu = \nu_0 + \beta t$, the relative optical phase evolves as $2\pi\beta \Delta t \cdot t$, where $\beta$ is the chirp parameter, $\Delta t$ the temporal misalignment and $t$ is the time.  In the presence of both large chirp and temporal misalignment, two wave-packets can oscillate in and out phase within the duration of their temporal overlap.

Time jitter of gain-switched lasers can be substantially reduced by injecting external photons to remove the uncertainty of the generation time of a spontaneous photon.\cite{chauchard1991,dong96}
We use a separate master laser for each slave laser to inject photons into their respective laser cavities. To maintain phase randomization the master lasers are also gain-switched across their respective lasing thresholds with 1~GHz square wave voltages producing $\sim$150~ps long pulses, each of which seeds a single slave pulse. Long master pulses relax the requirement for precise temporal alignment.
Gain-switching intervals of 500~ps are sufficient to allow the cavity photon field to decay and thus ensure each master laser pulse is triggered by spontaneous emission.\cite{yuan14}
With this seeding configuration we measure a reduction of the time jitter of the slave laser pulses by a factor of 3, see Table~I. Each slave laser produces an average output power of 360~$\mu$W.
In principle, the master laser can also have an adverse effect on the timing jitter performance of the slave laser, as it has considerable time jitter itself due to gain-switching.
We prevent a transfer of this jitter to the slave laser by choosing a low injection power of about 5~$\mu$W only in order for the electrical injection of 6.5~mA to dominate the carrier density build-up in the slave laser cavity.
With optical injection the time jitter of the slave laser pulses arises almost completely from the uncertainty in time of the carrier density overcoming the lasing threshold.

\begin{figure}[h]
\centering
\includegraphics[width=\columnwidth]{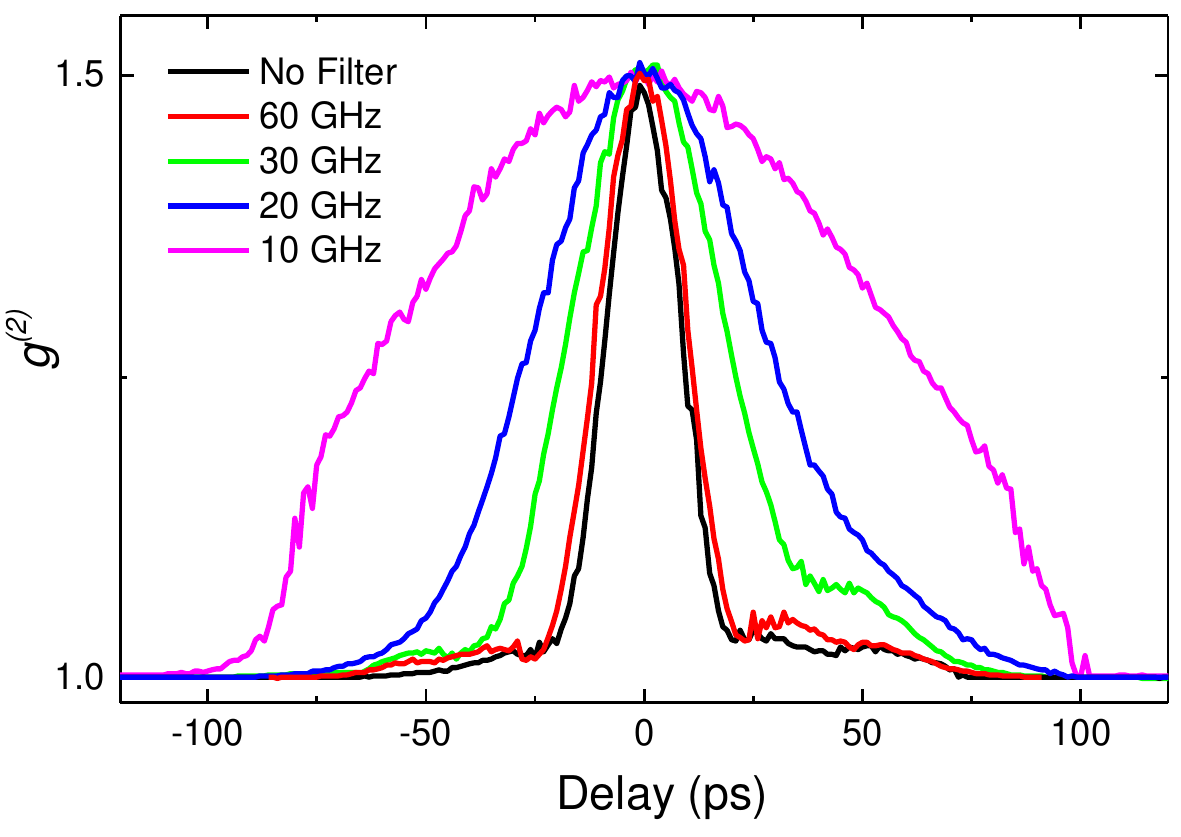}
\caption{Second-order correlation traces measured for different filtering bandwidths. A flat-top filter with tunable bandwidth is placed before the optical-to-electrical convertor in the setup of Fig.~1(a).}
\label{fig:fig3}
\newpage
\end{figure}

With the reduced time jitter we are able to obtain a drastic improvement of the second-order correlation value, see Fig. 2(c).  The measured peak value of 1.49 is very close to the theoretical maximum, suggesting an excellent mode match between the light sources.
Figure 1(c) shows an example oscilloscope trace which was used to extract the correlation value. Each data point is the result of the interference of a pulse from each independent light source. The random phase of the pulses causes the strong fluctuation of the measured intensity. Denser distribution around constructive ($I(t)=1)$ or destructive interference ($I(t) = 0$) suggests that the relative phase between the interfering pulses is uniformly distributed within [0,$2\pi$).\cite{yuan14}
We would like to stress that we achieve the high visibility despite the fact that the laser pulses are far from transform limited, as can be seen from their spectral shape in Fig. 2(b). Instead, we attribute the achieved visibility to synchronized frequency chirping.  The spectral alignment of the master lasers defines the same starting wavelength (shorter wavelength peak) for both light sources so that they can have a synchronized wavelength profile during their temporal overlap at the beam splitter.  Moreover, the optical injection reduces the overall spectral widths (full widths at 1/10 maximum) from 0.56~nm to 0.40~nm, and helps to resolve mode competition and achieve a side-mode suppression of over 30~dB, improving the match between the light sources.

Limiting the frequency chirp can further improve the mode match.  To demonstrate this, we record a number of second-order correlation traces after adding a tuneable spectral filter in front of the optical-to-electrical convertor.  As shown in Fig. 3, the dominant effect of spectral filtering is temporal broadening, which makes the delay alignment less critical.  Whenever a filtering bandwidth of less than 60~GHz is used, we become unable to resolve the difference among the peak correlation values. They all appear to have reached the theoretical limit, within the measurement fluctuation. 

\begin{figure}[b]
\centering
\includegraphics[width=\columnwidth]{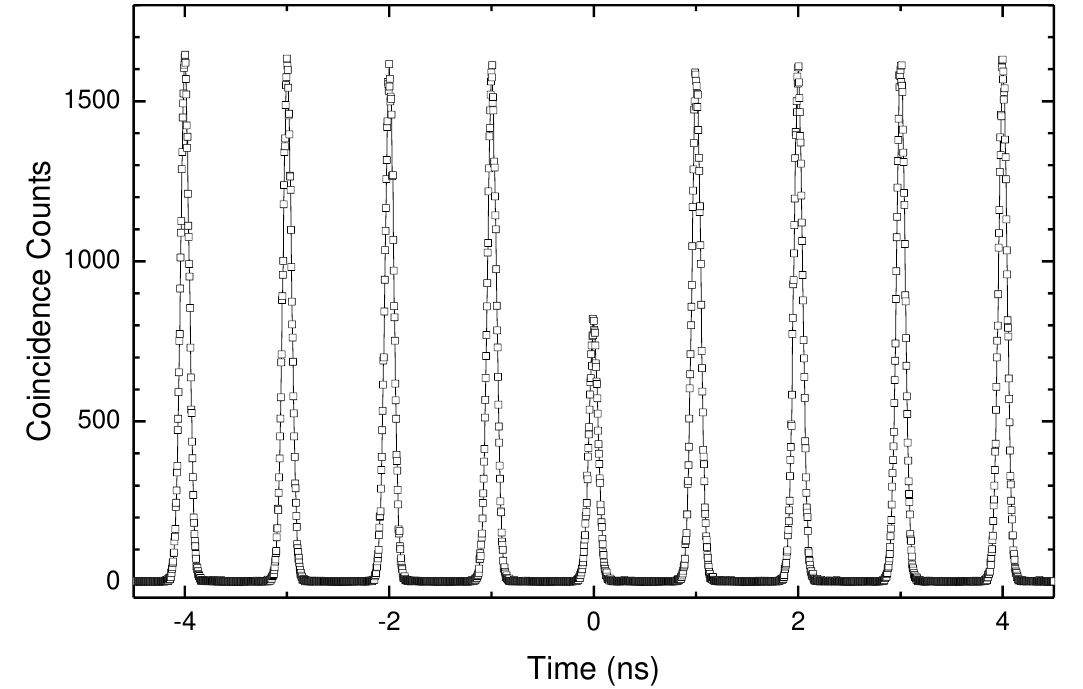}
\caption{Hong-Ou-Mandel interference coincidence counts using superconducting nanowire detectors, pulsed laser seeding and optical filters (11~GHz) for an acquisition time of 1 minute. The count rate at each detector is around 1~MHz.}
\label{fig:fig4}
\newpage
\end{figure}

To evaluate the mode match more precisely, we employ measurements in the single photon regime\cite{rarity05} to determine the HOM dip visibility $V_{HOM}$.  This measurement is useful also because laser pulses are usually attenuated to the single photon level in quantum information applications. We replace the optical-to-electrical convertor in Fig.~1(a) with a superconducting nanowire detector, and attach a second one to the other exit port of the 50/50 beam-splitter. These detectors have a single photon detection efficiency of 5\% at the wavelength of 1550~nm and their dark count rates are negligible as compared to the photon count rates as shown below. A spectral filter is applied to each light source resulting in a FWHM bandwidth of 11~GHz.  The light sources are attenuated to produce equal count rates ($\sim$1~MHz) at each detector. Time-tagging electronics are used to record the photon detection events. Figure 4 shows the HOM coincidence measurement result, which consists of a strong suppression at 0 time delay.  We compute the HOM dip visibility by comparing the counts in a 1 ns time bin around 0 delay with  the average of 10 adjacent 1~ns time bins. Without correcting for any noise background, we obtain a value of $V_{HOM} = 0.487 \pm 0.003$, where the error is the shot-noise statistical fluctuation due to the finite coincident count number.  By gating with a time window of 112~ps, equivalent to the FWHM width of each coincidence peak, the $V_{HOM}$ value improves to $0.499\pm0.004$.

\begin{table}[h!]
\centering
\caption{Performance comparison of various experiments to achieve independent, indistinguishable coherent light pulses. Perfectly indistinguishable coherent sources produce $V_{HOM}=0.5$.}
\label{my-label}
\begin{tabular}{c|c|c|c|c|c}
\hline \hline
                                        & \begin{tabular}[c]{@{}c@{}}clock\\ (MHz)\end{tabular} & \begin{tabular}[c]{@{}c@{}}width\\ (ps)\end{tabular} &          $V_{HOM}$             &           \begin{tabular}[c]{@{}c@{}} generation\\ method\end{tabular}   & \begin{tabular}[c]{@{}c@{}}require phase\\ randomization? \end{tabular}         \\ \hline \hline
da Silva \textit{et al.}\cite{silva13}            &          1    &         1500      &          0.478             &          pulse carving      & yes      \\ \hline

Rubenok \textit{et al.}\cite{rubenok13}&        2          &            500       &     0.47 $\pm$ 0.01          &             pulse carving       &yes         \\ \hline
Tang \textit{et al.}\cite{tang2014experimental} &         0.5       &         1000     &           0.475 $\pm$ 0.010     &           pulse carving     &yes       \\ \hline
Tang \textit{et. al.}\cite{tang16}     &        75     &   2500   &           0.46\footnote{Extracted from figure.}          &   internal modulation      & -   \\ \hline
\multicolumn{1}{c|}{\multirow{2}{*}{this work}} & \multicolumn{1}{c|}{\multirow{2}{*}{1000}}                & \multicolumn{1}{c|}{27}                                & \multicolumn{1}{c|}{  0.464 $\pm$ 0.003} & \multicolumn{1}{c}{gain-switching: PLS} & \multicolumn{1}{|c}{\multirow{2}{*}{no}} \\  \cline{3-5}
\multicolumn{1}{c|}{}                  & \multicolumn{1}{c|}{}                                 & \multicolumn{1}{c|}{45}                                & \multicolumn{1}{c|}{0.499 $\pm$ 0.004} & \multicolumn{1}{c|}{gain-switching: PLS + filtering}
\\ \hline \hline
\end{tabular}
\end{table}

For completeness, we measure the $V_{HOM}$ values by removing the spectral filtering and subsequently switching off the master lasers to disable the optical injection.
We apply a time gate equivalent to the FWHM of the corresponding coincidence peaks and obtain $V_{HOM}$ values of $0.463\pm0.003$ and $0.249\pm0.003$ for these cases, respectively. These results agree again with the second-order intensity correlation measurements (Fig.~2(c)), reconfirming that the PLS technique can substantially improve the interference visibility. It confirms also the validity of measuring the second-order correlation values as an effective method to evaluate the mode match between light sources.

The measured $V_{HOM}$ values can be reproduced with Monte Carlo simulations using experimentally measured parameters.  We compute the single-shot interference outcome $I$ between two incoming pulses of unity intensity using the equation \cite{yuan2014interference} (see Appendix for a derivation),
\begin{equation}
I =  1 + \cos(\Delta\varphi_0) \exp [-\frac {(\Delta t)^2} {8 \tau_p^2}(1+ 16 \beta^2 \tau_p^4)],
\label{eq:interference}
\end{equation}
\noindent where $\Delta \varphi_0$ is a random phase, $\Delta t$ is temporal misalignment, and $\tau_p$ is the Gaussian r.m.s. width of the input pulses. We determine the chirp coefficient $\beta$ using the relation\cite{agrawal02}
\begin{equation}
\Delta \omega^{(\beta)} = \Delta \omega^{(0)} \sqrt{1+16\beta^2\tau_p^4}
\end{equation}
\noindent between the measured bandwidth $\Delta \omega^{(\beta)}$ and the Fourier-transform limited bandwidth $\Delta \omega^{(0)}$ expected from the measured pulse width.  We use random numbers to draw a random phase ($\Delta \varphi_0$) uniformly distributed within [0,2$\pi$) and a temporal misalignment ($\Delta t$) assuming a Gaussian jitter profile for each single-shot interference event.  We build up a sample size of $10^6$ points to compute the second-order interference visibility using Eqs. 1 and 2.
For PLS with spectral filtering we use the following parameter set: $\Delta_\omega^{(\beta)} = 11$~GHz, $\tau_p = 19.1$~ps (pulse width: 45~ps) and jitter of 3.6~ps. The corresponding chirp coefficient is $\beta = 3.5 \times 10^{-4}$~ps$^{-2}$.
We run the simulation 100 times and obtain $V_{HOM} = 0.498 \pm 0.001$, which is in excellent agreement with the measured value of $0.499 \pm 0.004$  using single photon detectors.
For PLS without spectral filtering, the simulation produces $V_{HOM} = 0.463 \pm 0.001$, which again agrees with the single-photon measurement result. In this latter case, we use a bandwidth of 50~GHz and a pulse width of 30~ps, resulting in a
$\beta$ value of $5.0 \times 10^{-3}$~ps$^{-2}$, which is an order of magnitude higher than the filtered case.

Finally, we compare our results with other methods\cite{silva13,liu2013experimental,rubenok13,tang2014experimental,tang16,yuan2014interference} of  generating indistinguishable laser pulses for MDI-QKD applications. As summarised in Table~II, gain-switching produces the shortest pulses and can operate at much higher clock frequencies. With PLS and/or spectral filtering, this method exhibits high interference visibilities as revealed by the measured $g^{(2)}$ or $V_{HOM}$ values. These advantages have recently enabled the first megabit/second secure key rate in MDI-QKD.\cite{comandar2015quantum}  We acknowledge that use of short optical pulses places a stringent requirement on the precision of remote clock synchronization, which can become the dominant source for time jitter.  Fortunately, spectral filtering significantly relaxes the requirement for synchronization precision, as shown in Fig.~3.  For a decent visibility of $V_{HOM}=0.45$, comparable with those previously achieved using much longer pulses (Table I), the jitter tolerances are $\pm 12$ and $\pm 25$~ps for 20 and 10~GHz spectral filtering, respectively.  Such tolerances are readily achievable with available technologies.\cite{patel12prx}

In summary, we measure the interference in the macroscopic intensity regime for rapid evaluation of mode match between independent light sources and confirm the measurement results with both Hong-Ou-Mandel interference using single photon detectors and Monte-Carlo simulations.  We achieve near perfect mode match between independently seed, gain-switched laser diodes operating a clock frequency of 1~GHz.

We thank S. Kalliakos for setting up the superconducting single photon detector system. L. C. Comandar acknowledges personal support via the EPSRC funded CDT in Photonics System Development.


\section*{Appendix: I\lowercase{nterference between frequency chirped light pulses}}

\begin{figure}[b]
\centering
\includegraphics[width=8cm]{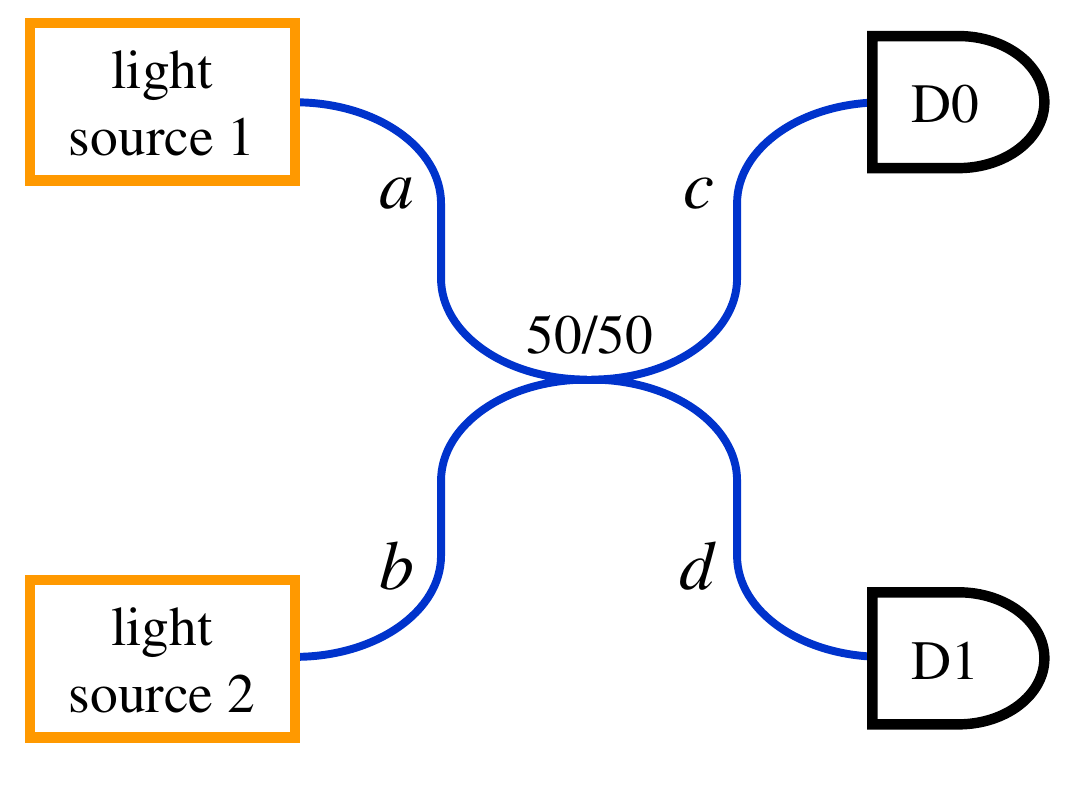}
\caption{Setup for interfering two independent light sources.}
\label{fig:figS1}
\newpage
\end{figure}

We consider the setup in Fig.~\ref{fig:figS1} and in particular the 50/50 beam splitter (BS) depicted therein. We label with $a$ and $b$ the BS's upper and lower input ports, respectively, and with $c$ and $d$ the BS's upper and lower output ports, respectively. The setup is such that the beams arriving at the BS from paths $a$ and $b$ have equal polarization. Therefore we omit the polarization vectors when writing the electric fields interfering at the BS and write them as:
\begin{eqnarray}
E_{a}(t)&=&\sqrt{I(t)}\exp\left[ i\left( \omega t+\beta t^{2}+\theta_{a}\right) \right] , \label{eq1} \\
E_{b}(t)&=&\sqrt{I(t)}\exp\left[ i\left( \omega t+\beta t^{2}+\theta_{b}\right) \right] , \label{eq2}
\end{eqnarray}
where
\begin{equation}
I(t)=\frac{I_{0}}{\tau_{p}\sqrt{2\pi}}\exp\left( -\frac{t^{2}}{2\tau_{p}^{2}}\right)\label{eq3}
\end{equation}
is their temporal shape, assumed to be Gaussian for simplicity. $I_{0}$ is the total intensity of each pulse, resulting from $\int_{-\infty}^{\infty}dtI(t)=I_{0}$; $\tau_{p}$ is the Gaussian r.m.s. width
of the pulses; $\omega$ is the central angular frequency of the lasers; $\beta$ is the chirping coefficient, with $\beta=0$ describing the zero-chirp condition. In writing Eqs.~\ref{eq1}--\ref{eq3}, we have also assumed that the interfering electric fields have well-defined electromagnetic phases and constant intensity. These conditions are well approximated by our laser sources, whereas do not hold, e.g., for thermal or bunched light.

By adding up the instantaneous electric fields interfering at the BS, we can obtain the output fields:
\begin{align}
E_{c}(t)& =\left[ E_{a}\left( t-\frac{\Delta t}{2}\right) +E_{b}\left( t+\frac{\Delta t}{2}\right) \right] /\sqrt{2}, \\
E_{d}(t)& =\left[ E_{a}\left( t-\frac{\Delta t}{2}\right) -E_{b}\left( t+\frac{\Delta t}{2}\right) \right] /\sqrt{2},
\end{align}
where $\Delta t$ is the relative time delay between the two pulses at the BS, depending on both systematic temporal misalignment and random time jitter of the light sources.  The factor $\sqrt{2}$ is due to the splitting of each pulse at the BS.

Let us now focus on the light seen by the detector on the $c$ path. The derivation for the light received by the other detector is analogous. For a small $I_{0}$, the intensity seen by the detector on the $c$ path is given by:
\begin{eqnarray}
I_{c}(t) &=&\left\vert E_{c}(t)\right\vert ^{2}  \notag \\
&=&\frac{I\left( t-\frac{\Delta t}{2}\right) +I\left( t+\frac{\Delta t}{2}%
\right) +\left[ E_{a}^{\ast }(t-\frac{\Delta t}{2})E_{b}(t+\frac{\Delta t}{2}%
)+c.c.\right] }{2}  \notag \\
&=&\frac{f+g\left[ e^{-i\left( \omega t-\omega \frac{\Delta t}{2}+\beta
t^{2}+\beta \frac{\Delta t^{2}}{4}-\beta t\Delta t+\theta
_{a}\right)}e^{i\left( \omega t+\omega \frac{\Delta t}{2}+\beta t^{2}+\beta
\frac{\Delta t^{2}}{4}+\beta t\Delta t+\theta _{b}\right) }+c.c.\right] }{2}
\notag \\
&=&\frac{f+g\left[ e^{i\left( \omega \Delta t+2\beta t\Delta
t+\theta_{b}-\theta _{a}\right) }+c.c.\right] }{2}  \notag \\
&=&\frac{f+2g\cos \left( \omega \Delta t+2\beta t\Delta t+\theta_{ba}\right)
}{2},\label{eq6}
\end{eqnarray}
where we have set:
\begin{eqnarray}
f &=&I\left( t-\frac{\Delta t}{2}\right) +I\left( t+\frac{\Delta t}{2}\right),
\\
g &=&\sqrt{I\left( t-\frac{\Delta t}{2}\right) I\left( t+\frac{\Delta t}{2}%
\right) } ,\\
\theta _{ba} &=&\theta _{b}-\theta _{a}.
\end{eqnarray}
However, the detector will not see the above instantaneous intensity, because of its own finite bandwidth $\nu =1/T$ which averages out all the terms oscillating faster than $\nu $.
Therefore, we average each of the above equations over the finite time $T$. To do that, we rewrite the $\cos$ term in Eq.~\ref{eq6} as follows:
\begin{eqnarray}
\cos \left( \omega \Delta t+2\beta t\Delta t+\theta _{ba}\right)  &=&\cos
\left( 2\beta t\Delta t\right) \cos \left( \omega \Delta t+\theta
_{ba}\right)  \notag \\
&&-\sin \left( 2\beta t\Delta t\right) \sin \left( \omega \Delta t+\theta
_{ba}\right) .
\end{eqnarray}
The term containing $\sin \left( 2\beta t\Delta t\right) $ will give zero
upon integration on the symmetric interval $t\in \lbrack -T/2,T/2]$. The
remaining terms are integrated considering the explicit Gaussian form of the
pulses, Eq.~\ref{eq3}:
\begin{eqnarray}
\left\langle I_{c}(t)\right\rangle _{T} &=&\frac{1}{2}\int_{-T/2}^{T/2}dt%
\left\{ f+2g\cos \left( 2\beta t\Delta t\right) \cos \left( \omega \Delta
t+\theta _{ba}\right) \right\}   \notag \\
&=&\frac{I_{0}}{2}+\frac{I_{0}}{2}+\frac{I_{0}\cos \left( \omega \Delta
t+\theta _{ba}\right) }{\tau _{p}\sqrt{2\pi }}\times   \notag \\
&\times &\int_{-T/2}^{T/2}dt\exp \left[ -\frac{\left( t-\frac{\Delta t}{2}%
\right) ^{2}+\left( t+\frac{\Delta t}{2}\right) ^{2}}{4\tau _{p}^{2}}\right]
\cos \left( 2\beta t\Delta t\right).
\end{eqnarray}
Because the detector's response time is larger than the pulses' width, we are allowed to solve the above integral in the limit $T\rightarrow \infty $, thus obtaining the following solution:
\begin{equation}
\left\langle I_{c}(t)\right\rangle _{T}=I_{0}\left\{ 1+\cos \left( \omega
\Delta t+\theta _{ba}\right) \exp \left[ -\frac{\Delta t^{2}}{8\tau _{p}^{2}}%
\left( 1+16\beta ^{2}\tau _{p}^{4}\right) \right] \right\}.
\end{equation}
The above equation has been reported as Eq.~3 in the main text. A similar expression can be derived for the light emerging from the port $d$ of the BS:
\begin{equation}
\left\langle I_{d}(t)\right\rangle _{T}=I_{0}\left\{ 1-\cos \left( \omega
\Delta t+\theta _{ba}\right) \exp \left[ -\frac{\Delta t^{2}}{8\tau _{p}^{2}}
\left( 1+16\beta ^{2}\tau _{p}^{4}\right) \right] \right\}.
\end{equation}

%


\newpage

\end{document}